# Near-surface Characterization Using a Roadside Distributed Acoustic Sensing Array

Siyuan Yuan*[1], Ariel Lellouch[1], Robert G. Clapp[1], Biondo Biondi[1]

[1]Stanford University, Stanford, CA
*corresponding author

**Abstract**

Thanks to the broadband nature of the Distributed Acoustic Sensing (DAS) measurement, a roadside section of the Stanford DAS-2 array can record seismic signals from various sources. For example, it measures the earth's quasi-static distortion caused by the weight of cars (<0.8 Hz), and Rayleigh waves induced by earthquakes (<3 Hz) and by dynamic car-road interactions (3-20 Hz). We directly utilize the excited surface waves for shallow shear-wave velocity inversion. Rayleigh waves induced by passing cars have a consistent fundamental mode and a noisier first mode. By stacking dispersion images of 33 passing cars, we obtain stable dispersion images. The frequency range of the fundamental mode can be extended by adding the low-frequency earthquake-induced Rayleigh waves. Thanks to the extended frequency range, we can achieve better depth coverage and resolution for shear-wave velocity inversion.  In order to assure clear separation from Love waves and aligning apparent velocity with phase velocity, we choose an earthquake that is approximately in line with the array. The inverted models match those obtained by a conventional geophone survey performed by a geotechnical service company contracted by Stanford University using active sources from the surface until about 50 meters. In order to automate the $V_S$ inversion process, we introduce a new objective function that avoids manual dispersion curve picking. We construct a 2-D $V_S$ profile by performing independent 1-D inversions at multiple locations along the fiber. From the low-frequency quasi-static distortion recordings, we invert for a single Poisson's ratio at each location along the fiber. We observe spatial heterogeneity of both $V_S$ and Poisson's ratio profiles. Our approach is dramatically cheaper than ambient field interferometry and reliable estimates can be obtained more frequently as no lengthy cross-correlations are required.

**Introduction**

Distributed Acoustic Sensing (DAS) transforms fiber-optic cables into a long-offset virtual sensor array with a high temporal sampling rate and meter-scale channel spacing. By leveraging standard existing fibers, widely installed for telecommunication, DAS could enable a large-scale seismic monitoring in urban and suburban areas thanks to its cost-efficiency, easy maintenance, and low imprint (Martin et al., 2017a; Biondi et al., 2020). This paper focuses on the application of a roadside DAS array at Stanford to shallow $V_S$ and Poisson's ratio characterization.

Near-surface characterization with DAS has been an active research topic during the last years. Previous studies have successfully used ambient noise interferometry, which is cost-effective, non-intrusive, and does not require active surveys (Martin et al., 2016; Dou et al., 2017; Zeng et al., 2017; Martin et al., 2017b). However, there are many challenges associated with this approach. It assumes independent, randomly distributed, and diffusive ambient noise sources. To satisfy this assumption, non-trivial preprocessing steps have to be applied to remove coherent noise (e.g. traffic, earthquakes, and construction noise). Additionally, the method requires long-term recording to converge. Therefore, subsurface changes, such as the formation of sinkholes, subsidence, and underground water leakage or accumulation, can only be estimated at a relative long-time scale, of the order of days. In addition, higher frequencies are often attenuated as the ambient sources are distant, and the cross-correlated signal is of low frequency. Additionally, downhole DAS recordings of body waves are used to estimate the velocity structure from direct arrivals (Lellouch et al., 2019b), and detect earthquakes (e.g. Lindsey et al., 2017; Lellouch et al., 2019a). However, it is challenging to estimate the velocity profile in depth using body waves recorded by a shallow horizontal fiber. It is practically impossible to separate velocity effects in the shallow subsurface from deeper parts of the subsurface. In addition, the wavefront's apparent velocity strongly depends on the deep velocity structure as well as the source depth, which are only approximately known.

In this study, we use surface waves passively recorded and directly generated by cars and earthquakes. There is no interferometric component involved. The recording is close to the source, and high frequencies can be used as they are not severely attenuated. We use the frequency-wavenumber (FK) method (Capon, 1969; Louie, 2001; Park, 2005) to compute dispersion images and, subsequently, the Multichannel Analysis of Surface Wave (MASW) method (Park et al. 1999) for shear-wave velocity profile estimation. We use a global optimization technique, the Particle Swarm Optimization (PSO), to invert for the $V_S$ profile (Eberhart and Kennedy, 1995; Luu et al., 2018). PSO has been shown to have good convergence rates and straightforward parameter tuning for $V_S$ inversion (Shragge and Yang, 2019). However, conventional MASW requires picking dispersion curves. In order to automate the process, we also suggest an alternative, picking-free objective function. We invert $V_S$ using this objective function as well as the conventional picks-based inversion. We compare both profiles with an independent result from an active survey, Spectral Analysis of Surface Waves (SASW, Stokoe et al. 1994), conducted in 2009 along a survey line located close to the DAS fiber (Wong et al., 2011).

DAS recordings of quasi-static distortions can be used to determine subsurface properties with controlled driving of a known vehicle (Jousset et al. 2018). In this study, we introduce an elegant way to map the fiber location, which is often unknown in urban areas due to generally poor fiber mapping quality, by dedicatedly driving a car across the fiber. Additionally, we compute Poisson's ratio from regular, passive traffic recording. It does not depend on the weight of the car, which is unknown in this case.

**Data Acquisition**

The Stanford DAS-2 experiment, based on an OptaSense ODH-3 interrogator, has been continuously recording since December 10, 2019. The sampling rate is 250 samples per second. The gauge length is 20 m. There are, in total, 1250 channels with a spacing of 8.16 m. About 350 channels are located along the relatively straight Sandhill road section spanning from Stanford hospital (Channel #400) to the Stanford Linear

Accelerator (Channel #750). Figure 1 shows the location of the Sandhill section within the Bay area.

## Recorded data

The Portola Valley earthquake occurred on March 18, 2020 at 00:11:32 AM local time. It is approximately in line with the section of the array we use shown in Figure 1(a). Due to the directive nature of the DAS measurement, Rayleigh waves are expected to dominate the surface wave energy. We focus on a subset of the Sandhill section, from channel #400 to channel #470, spanning 563 m as shown in Figure 2 (a). We use this full section for earthquake-based analysis, as we target lower frequencies. Recorded traffic, on the contrary, generates surface waves with much higher frequencies, and we thus use a shorter subsection for its analysis. Figure 2 (b) shows DAS recordings of the Portola earthquake, acquired in the subsection of the fiber colored in red. The average Power Spectral Density (PSD) between the two dashed yellow lines is computed and will be compared with the spectrum of a car recording. For surface wave dispersion analysis, we use the recordings between 12 s and 15 s (red box).

In Figure 2 (c), we show the recording of a car driving along the fiber, at midnight, after we apply a bandpass filter of 0.2 Hz to 30 Hz. The strong low-frequency energy is due to the quasi-static deformation caused by the weight of the car. Thanks to the quasi-static signal, we can accurately count cars and track their positions along the fiber to better understand city-scale human mobility (Huot et al., 2019; Lindsey et al., 2020; Wang et al., 2020). The car's driving speed can be measured by computing the slope of the quasi-static data. If we apply a 0.8 Hz to 30 Hz bandpass filter, the quasi-static component is filtered out and surface waves, excited by the car and propagating along the fiber, are visible, as shown in Figure 2 (d). Data within the green box are recorded by the green section shown in Figure 1 (a) and will be used for dispersion analysis. The average PSD of recordings within the two dashed yellow lines is computed and compared with the earthquake spectrum in Figure 2 (e). The earthquake has an approximately flat spectrum at frequencies between ~1.2 and ~10 Hz, whereas the car-induced surface waves have an approximately flat spectrum at frequencies between ~3 and ~16 Hz. However, we can improve signal quality in the traffic frequency band can by stacking multiple cars. We eventually combine the two types of sources, earthquake (< ~3 Hz) and cars (> ~3Hz), to broaden the spectrum that is subsequently used for dispersion analysis.

**Dispersion and $V_s$ inversion**

Figure 3 (a) shows a dispersion image computed for a single car using the FK method. Its fundamental mode ranges from 5 Hz to 15 Hz and is clearly visible. The first mode is noisier yet visible in the range of 10-15 Hz. In Figure 3 (b), we show a dispersion image obtained by stacking 33 different cars traveling along the analyzed fiber section. Much smoother dispersion images can be obtained by this procedure. In addition, the first mode is now clearly visible. Figure 3 (c) shows the dispersion image of the Portola Valley earthquake, which has dominating low-frequency energy. Figure 4(d) shows a combination of the two dispersion images – stacked cars for frequencies above 2.7 Hz, and the earthquake for frequencies below 2.7 Hz. They can be combined, and the joint fundamental mode has a more complete frequency band, ranging from 1.3 Hz to 15 Hz. The first mode after stacking is stable and can be reliably used between 11 Hz and 18 Hz.

We apply PSO (Luu et al., 2018) to invert for $V_S$ model, using 100 iterations and 120 particles. We invert for 10 layers with variable width and weak constraints on the S-wave velocity. Traditionally, surface wave analysis requires picking of dispersion curves. The objective function is then

$$\min ||\boldsymbol{d(f)} - \boldsymbol{d(f)}^*||_2^2,$$

where $\boldsymbol{d}$ and $\boldsymbol{d}^*$ are the picked and simulated dispersion curves, representing phase velocity as function of frequency $f$, respectively. To automate the workflow, we introduce an alternative objective function for surface wave inversion. It is adequate for global optimization methods. Instead of fitting the picks, we compute a weighted sum of the dispersion spectrum along each simulated dispersion curve $\boldsymbol{d}^*$, and we maximize the weighted sum,

$$max \int_{\boldsymbol{d}^*} D(f, c)w(f)df,$$

where D is the dispersion image, representing amplitudes of frequency and phase velocity pairs, and $w(f)$ is the weight of frequency $f$. In our case, the fundamental mode is unequivocal from 1.2 Hz to 2.7 Hz and from 4 Hz to 15 Hz. Therefore, we set the weight at 1 for this frequency range. For the fundamental mode between 2.7 and 4 Hz, and the first mode between 11 and 18 Hz, we set the weight at 0.5 as the signal is weaker. This is an empirical choice only. Other frequencies contain no clear energy associated with either mode and are thus assigned a zero weight. In Figure 4(a), we show the dispersion images, with picks of the fundamental and first modes overlaid in white dots. The picking-based inversion yields the dark green (cars only) and dashed black (cars and earthquake picks) curves. For both cases, we fit the picked dispersion curves well, indicating that the inverted $V_S$ modes are reliable. Using the picks-free objective function (all other inversion parameters are left unchanged), we obtain the magenta curve in Figure 4. It also fits the picked dispersion curves well, even though the picks are not involved in the procedure. Figure 4 (b) shows the matching inverted models, represented with matching colors. Using the joint stacked cars and earthquake dispersion image, our results match the 2009 active-sources survey (Wong et al., 2011) well for the top 50 m. A velocity inversion at a depth of 15 m is visible in all models. Interestingly, our results show that the velocity increases gradually, indicating a compaction effect, whereas the results of the survey in 2009 show a constant $V_S$ from ~30 to 75 m. Active sources are unlikely to have energy in the low frequency range excited by the earthquake. The SASW survey may thus not properly resolve the structure at depths of more than ~50 m. The seismic wavelength induced by the earthquake can be as long as ~400 m at 1.5 Hz. For the cars, the wavelength is shorter than ~200 m for frequencies above 2.7 Hz. The maximum investigation depth using surface waves is about half the longest wavelength (Park et al., 2002). Therefore, with the earthquake we can retrieve depths down to ~200 m, whereas using only the cars, results at depths below ~100 m are unreliable. The model estimated by the automatic procedure agrees well with the one estimated using picking. Thus, a fully automated $V_S$ monitoring from field recording without manual intervention is feasible, once we've determined the quality of the dispersion images at different frequency bands to assign their weights.

We follow the picking-free workflow at eight other locations along the fiber. These locations are 81.6 m (10 channels) apart (Figure 5), and fiber subsections with 13 channels centered around them are used. Figure 6 shows the combined dispersion images of the stacked cars and the earthquake for each location. The two black curves in each panel represent simulated dispersion curves for the first two modes using the inverted models, obtained with the picking-free inversion, shown in Figure 7. Overall, we observe structural coherency, which is important for interpretation.  Except for the locations from channel 472 to 492, there is a velocity inversion at ~15 m. The top 25 m, which are possibly a deposit layer, have velocities below 500 m/s. Layers generally below 45 m and with a velocity of above 600 m/s can be interpreted as stiffer rock.

**Fiber location and Poisson's ratio estimation**

Quasi-static deformation, which is the low-frequency strain induced by the weight of passing cars, also be recorded by DAS. As the cars' velocity is much slower than Rayleigh waves, we can use the Flamant-Boussinesq theory (Fung, 1965), which models the static deformation of a point load on the ground. The displacement along the fiber, $u_x(x)$, is given by

$$u_x(x) = \frac{F}{4\pi\rho v_p^2}\frac{2v-2}{2v-1}\left(\frac{zx}{r^3} - \frac{x}{r^2+zr} + \frac{2xv}{r^2+zr}\right)$$

Where $x$ denotes the fiber axis, F is the gravitational force the car applies on the ground, $\rho$ is the density, $v_p$ is the compressional wave velocity, $v$ is the Poisson's ratio, $z$ is the depth of the fiber, $x$ is the distance from the source to the receiver along the fiber, and $r$ is the 3-D distance from the source to receiver. This equation can be used directly for the estimation of elastic parameters (Jousset et al., 2018). However, we can see from this equation that F, $\rho$ and $v_p$ affect only the amplitude of the displacement signal, whereas $v$ and the relative position of the car to the fiber determine its shape. The same property is true when converted to strain measurements of DAS. We thus utilize solely the recorded signal's shape to locate the fiber channels in relation to the road and to estimate Poisson's ratio, in two separate steps. We can thus avoid handling a quantitative analysis of the fiber's optical response, coupling, etc. For mapping purposes, we drove a car crossing the fiber at two locations (~3m apart) at the Clark way-Sandhill road intersection, as shown in Figure 8. The car speeds for the two passes are 3.4 m/s and 2 m/s, respectively. Figure 8 (a) and (c) show the DAS recordings of the two passes. We model the four wheels of the car as four equal concentration forces and simulate recorded strain induced by the moving cars using their known speeds. We perform a grid search for each car pass, with the goal of matching recorded data, by looping over relative distances (dx1, dx2) from the car to channel 415, $v$ (0.1 ~ 0.48), and z (0.2 ~ 3 m). The objective function is a normalized cross-correlation. The parameters of the best fit for the first pass are dx1=1.12 m, z=0.6 m and $v = 0.33$, and for the second pass are dx2=-1.88 m, z=0.6 m, and $v = 0.35$. The simulated data using the best fit parameters are shown in Figure 8 (b) and (d). The normalized cross-correlations are, respectively, 0.94 and 0.96. Imprints of back and front wheels are visible in both simulated and real data. Both have reverse polarity at the wheels between the passes, because of the sidelobes in the strain response to the passing car. These results agree with the fact that the two passes are 3 m apart. The two independent inversions also yield the same fiber depth of 0.6 m. Additionally, their Poisson's ratio estimates are very close. The average estimated $\nu$ is 0.34, which is consistent with asphalt concrete (typically, $\nu \approx 0.35$). With the car position known at each time, we estimate the distance between the fiber to the first lane of the Sandhill road to be 6.1 m.

Assuming the fiber depth of 0.6 m and fiber-road offset of 6.1 m stay constant, we invert for the Poisson's ratio using passing cars along the ~1 km section of the Sandhill road. At each location, we estimate a single value. We manually choose, from regular traffic, 5 cars per day for March 1st, 10th, 15th, 20th, 25th and 30th 2020. All of these chosen cars passed the entire section of interest. The recordings we use for inversion are three seconds long and with a 60 m offset on both sides of the moving car. We compute the car speed, needed for inversion, using a local slant-stack measurement. We loop through Poisson's ratio values of 0.1 to 0.5 to maximize the normalized cross-correlation between the simulated and real data. For each car, we find the Poisson's ratio at different locations, and subsequently average over five estimates per day. Daily averages for the six days in March are shown in Figure 9. We can see that these estimates show consistent spatial variations. Poisson's ratio is on average ~0.44. Since the majority of the fiber section is buried under grass and March is a wet season in the Bay area, the Poisson's ratio estimates suggest the fiber is installed in saturated clay, with a typical Poisson's ratio between 0.40 and 0.49. We can see the Poisson's ratio drops at channels 422, 472 and 512. All of these channels are centered at intersections, where fiber is buried under the road, which has a lower Poisson's ratio than clay. The harder pavement could thus decrease the Poisson's ratio estimates at those locations since we measure an average, effective property.

**Discussion and Conclusions**

We introduce a new approach to ambient noise interferometry for passive near-surface property estimation using a road-side DAS array. Instead of extracting surface waves by cross-correlating long periods of ambient noise records, we leverage Rayleigh waves directly excited by cars and a local, favorably aligned, earthquake. We stack the signal generated by many cars passing the same section of the fiber, thus significantly improving dispersion images. We show that the earthquake and cars generate surface waves with complementary frequency bands. Thus, we can merge the dispersion images of the two types of sources to broaden the frequency band, which leads to better depth coverage and resolution for the $V_S$ inversion. We introduce a new objective function for $V_S$ inversion that avoids dispersion curve picking. We compare the results with those obtained by fitting manual picks with and without low frequencies from an earthquake. All of the inverted $V_S$ models agree well with an active-sources geophone-based SASW survey conducted at a nearby area in 2009. In addition, we obtain more realistic results at depth greater than 50 m, thanks to the low frequencies, hard to actively excite, induced by the earthquake. Unlike interferometry, this method does not require various preprocessing steps (spectral whitening and coherent signal removal, for example). Computation is cheaper, because we do not need to cross-correlate long-term recordings. Additionally, our method can quickly deliver results, as we have seen that as few as 33 cars passing the road can yield a stable and smooth dispersion image. Earthquakes improve the inversion results but are not necessary for near-surface (<100 m depth) models, which are the area of interest for most engineering purposes. This approach is practically interesting because it can detect changes in close to real time at no additional cost, given reasonable traffic activity. It can thus be applied to the detection of sinkholes, underground water leakage or accumulation, and subsidence. We independently apply the approach at nine locations along the fiber and construct a 2-D $V_S$ model.

Furthermore, we show that from quasi-static data, we can invert channel locations, fiber depth and Poisson's ratio. Locations of the fiber-optic cable can be challenging to recover, especially in urban settings, where complex network of different types of underground utility grows rapidly. Our method of locating channels is a non-intrusive approach, that could be promising to satisfy the need of swiftly and precisely detecting these utilities in a noisy urban environment. Once the fiber depth and location is known, we build a 1-D Poisson's ratio profiles along the fiber for different days in March 2020. These profiles are consistent with our geospatial knowledge of near-surface properties, mostly of grass versus road intersections.

**Acknowledgements**

This research was financially supported by the affiliates of the Stanford Exploration Project. The IU was loaned to us by OptaSense Inc. We thank Martin Karrenbach, Victor Yartsev, and Lisa LaFlame from Optasense, as well as the Stanford ITS fiber team, and in particular Erich Snow, for crucial help with the Stanford DAS-2 experiment. We would like also the Stanford School of Earth IT team for hosting the interrogator in the Scholl computer room.

## LIST OF FIGURES

Figure 1 - (a) The location of the Sandhill section (black line segment) of the Stanford DAS-2 array in Bay area. Red point shows the location of a Portola Valley earthquake (magnitude 1.1, depth 8.2 km). It is approximately in-line with the DAS array and located ~9.3 km away in horizontal distance. (b) A zoomed-in view of the Sandhill section with channel numbers and distance along the fiber labeled.

Figure 2 - (a) A zoomed-in view of the study area of the Stanford DAS-2 array. The red segment is used for earthquake analysis and the green for cars. The blue point indicates the location of the geophone SASW survey conducted in 2009. Channel #470 is chosen as the distance origin for displaying data in (b), (c) and (d). (b) A Portola Valley Earthquake of magnitude 1.1 recorded by the subsection of DAS corlored in red in Figure 1. The origin of the distance axis is ~9.3 km away from the event. Data within the red box is the surface wave used for dispersion analysis. Data within the two dashed lines are used to compute an average PSD spectrum. (c) DAS recording of a car driving along the red section shown in Figure 1 with a bandpass 0.2 Hz to 30 Hz; (d) Same data as (c), but bandpassed between 0.8Hz and 30Hz. Data within the green box is used for dispersion analysis. Data within the two dashed lines are used to compute an average PSD spectrum. Orange and blue curves represent average PSD spectra of recordings within the yellow dashed lines in (b) and (d), respectively. The earthquake dominates frequencies between ~1.2 and ~10 Hz. The car dominates frequencies between ~3 and ~16 Hz.

Figure3 - (a) Dispersion image for a single car passing the green subsection. The fundamental (magenta ellipse) and first (black ellipse) modes are visible. (b) The dispersion image of stack of 33 cars are much smoother than (a) with clearer fundamental (magenta ellipse) and first (black ellipse) modes. (c) Portola Valley earthquake; (d) Combination of the car average and the earthquake. The combined spectrum is broad, smooth, and dispersion curves can be easily picked automatically.

Figure 4 - (a) Picked (white dots) and predicted (dark green, dashed black and solid black curves) dispersion curves using estimated $V_S$ models shown in (b). To the left and right of the vertical gray line at 2.7 Hz are picks from the earthquake and the 33-car stack dispersion images, respectively. The dashed black and the dark green curves are fitted using picks with and without the earthquake combined, respectively. The solid black curve is fitted using the automatic approach, with no picks involved.  Fits are very good for all cases, and the first mode stabilizes the inversion. (b) Inverted $V_S$ models via PSO. The red curve represents a $V_S$  model obtained from a SASW survey in 2009. The dashed black and the dark green models are inverted using picks with and without earthquake combined. The solid black curve is obtained without any picks. In the shallow section, inverted models from DAS data agree very well with the SASW survey, and both show the velocity inversion at a depth of ~15 m. $V_S$ inverted from car pickings (dark green) can be unreliable below ~100 m, because of the lack of low frequencies below 2.7 Hz.

Figure 5 - Satellite map of the Sandhill fiber (red line). Locations where both $V_S$ and Poisson's ratio inversion are performed are shown with orange triangles, whereas, Poisson's ratio-only locations are shown with green triangles. Magenta lines indicate intersections. The white ellipse circles out the area where we conducted fiber location experiments by driving our car orthogonal to the fiber.

Figure 6 - Dispersion images for eight locations (indicated in Figure 5) along the fiber. Frequencies below 2.7 Hz are from the Portola Valley earthquake. Above 2.7 Hz are from 33-car stack dispersion images. The proposed approach that does not require dispersion curve pickings are used for $V_S$ inversion. The simulated dispersion curves using the inverted model are overlaid on the dispersion images using black curves. Inverted Vs models are shown in Figure 7.

Figure 7 - Inverted Vs models. The profile of the channel 422 comes from the solid black curves in Figure 4 (b). Results of the other channels are inverted from dispersion images shown in Figure 6. Except channel 472 and 492, we have velocity inversion at ~15 m. The top 25 m possible to be a deposit layer tends to have velocity below 500 m/s. Layers below 45 m with $V_S$ above 600 m/s could be interpreted as hard rock.

Figure 8 - (Top) A zoomed-in map view of the intersection for our car-passing experiment. Car speeds for the two passes are respectively 3.4 m/s and 2 m/s. (Bottom) (a) and (c) show the data produced by the pass #1 and #2, respectively.

Figure (b) and (d) show the simulated data using the parameters inverted through grid searches. The normalized cross-correlations are 0.94 between (a) and (b), and 0.96 between (c) and (d). The parameters of the best fit for the first pass are dx1=1.12 m, z=0.6 m and ν=0.33. For the second pass are dx2=-1.88 m, z=0.6 m, and ν=0.35. These results agree with the fact that the two passes are 3 m apart. The two inversions lead to a same fiber depth, 0.6 m. their Poisson's ratio estimates are close. The average estimated ν is 0.34 agreeing with a pavement material, Asphalt Concrete of a typical Poisson's ratio of 0.35 (±). Imprints of front and back wheels are pointed out with arrows in (b) and (d).

Figure 9 - Daily averages of the Poisson's ratio estimates for the six days in March 2020. Poisson's ratios on average are ~0.44, indicating the shallow subsurface to be saturated clay. see Poisson's ratio drops at channels 422, 472 and 512 are likely related to their close locations to intersections, where fiber is buried under the road.

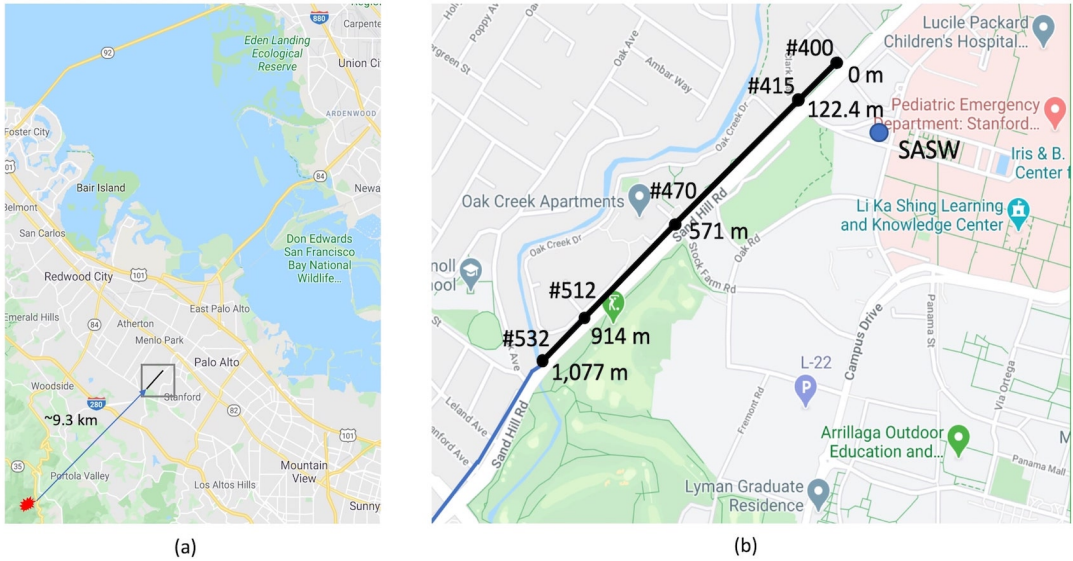

Figure 1: (a) The location of the Sandhill section (black line segment) of the Stanford DAS-2 array in Bay area. Red point shows the location of a Portola Valley earthquake (magnitude 1.1, depth 8.2 km). It is approximately in-line with the DAS array and located ~9.3 km away in horizontal distance. (b) A zoomed-in view of the Sandhill section with channel numbers and distance along the fiber labeled.

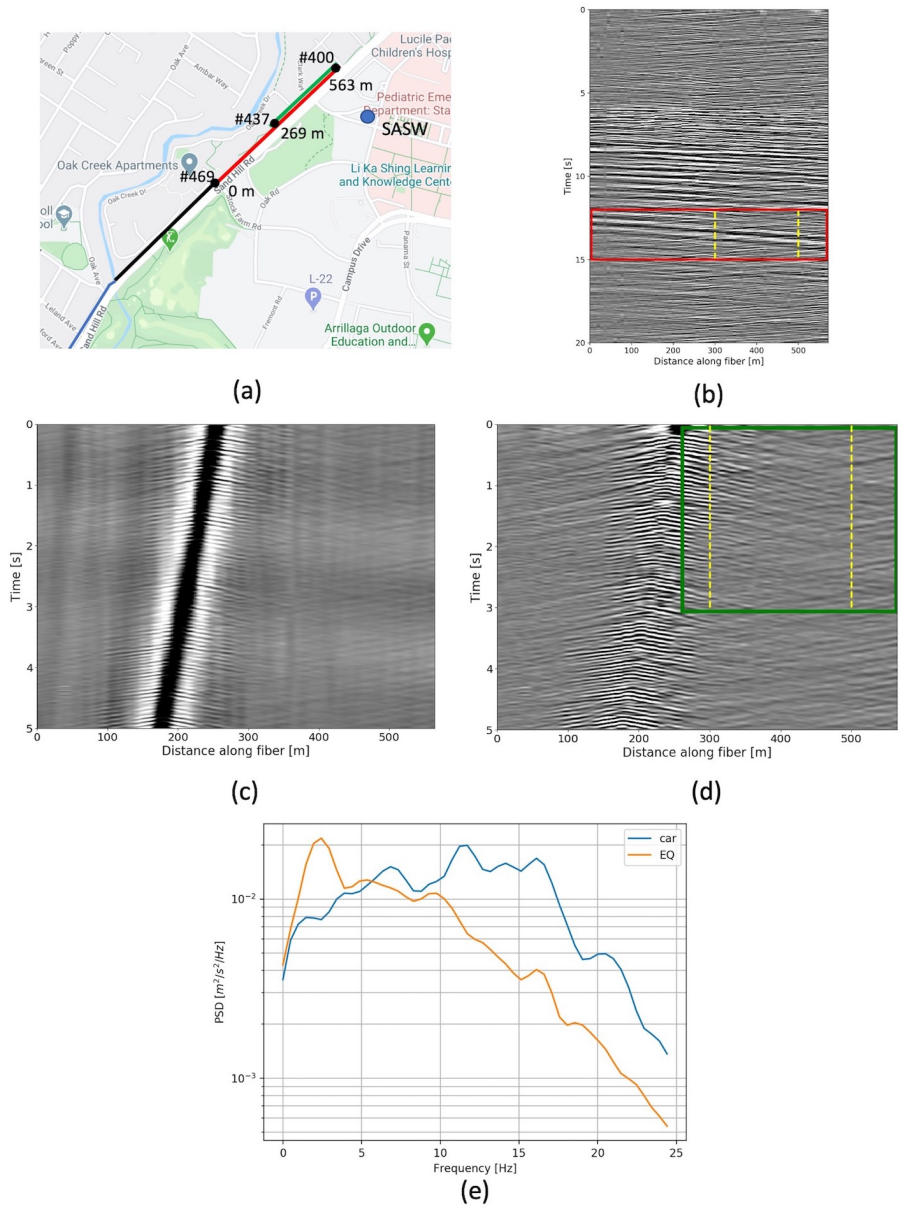

Figure 2: (a) A zoomed-in view of the study area of the Stanford DAS-2 array. The red segment is used for earthquake analysis and the green for cars. The blue point indicates the location of the geophone SASW survey conducted in 2009. Channel #470 is chosen as the distance origin for displaying data in (b), (c) and (d). (b) A Portola Valley Earthquake of magnitude 1.1 recorded by the subsection of DAS corlored in red in Figure 1. The origin of the distance axis is ~9.3 km away from the event. Data within the red box is the surface wave used for dispersion analysis. Data within the two dashed lines are used to compute an average PSD spectrum. (c) DAS recording of a car driving along the red section shown in Figure 1 with a bandpass 0.2 Hz to 30 Hz; (d) Same data as (c), but bandpassed between 0.8Hz and 30Hz. Data within the green box is used for dispersion analysis. Data within the two dashed lines are used to compute an average PSD spectrum. Orange and blue curves represent average PSD

spectra of recordings within the yellow dashed lines in (b) and (d), respectively. The earthquake dominates frequencies between ~1.2 and ~10 Hz. The car dominates frequencies between ~3 and ~16 Hz.

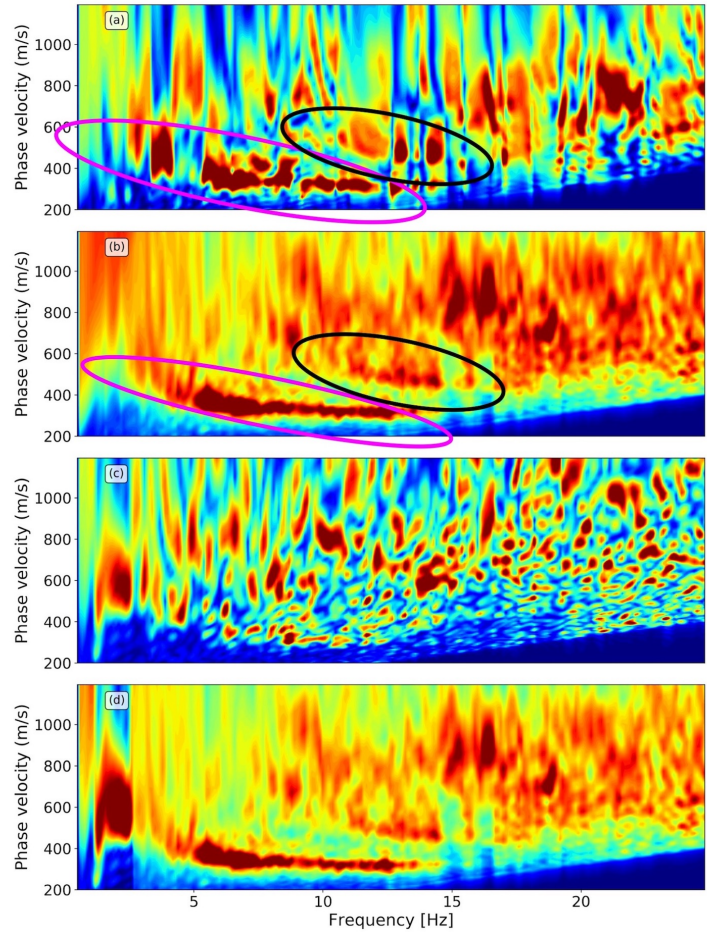

Figure 3: (a) Dispersion image for a single car passing the green subsection. The fundamental (magenta ellipse) and first (black ellipse) modes are visible. (b) The dispersion image of stack of 33 cars are much smoother than (a) with clearer fundamental (magenta ellipse) and first (black ellipse) modes. (c) Portola Valley earthquake; (d) Combination of the car average and the earthquake. The combined spectrum is broad, smooth, and dispersion curves can be easily picked automatically.

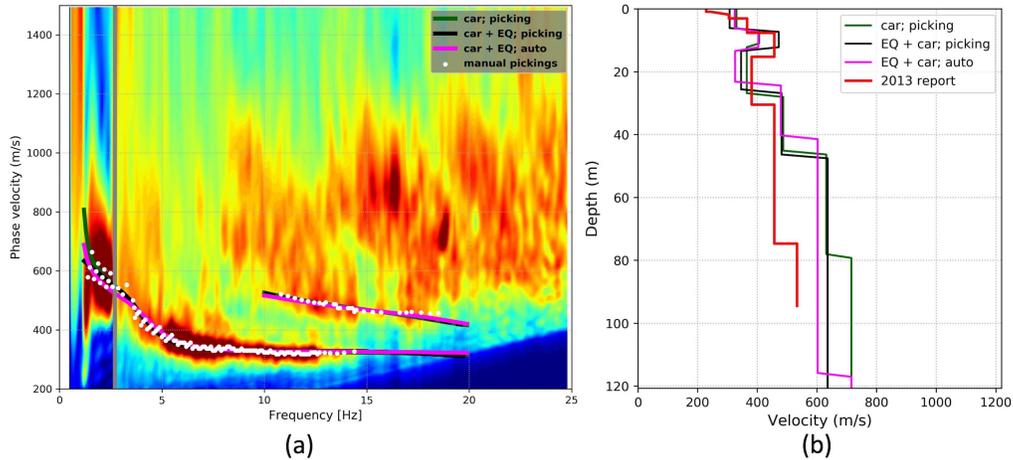

(a)                                                        (b)

Figure 4: (a) Picked (white dots) and predicted (dark green, dashed black and solid black curves) dispersion curves using estimated $V_S$ models shown in (b). To the left and right of the vertical gray line at 2.7 Hz are picks from the earthquake and the 33-car stack dispersion images, respectively. The dashed black and the dark green curves are fitted using picks with and without the earthquake combined, respectively. The solid black curve is fitted using the automatic approach, with no picks involved. Fits are very good for all cases, and the first mode stabilizes the inversion. (b) Inverted $V_S$ models via PSO. The red curve represents a Vs model obtained from a SASW survey in 2009. The dashed black and the dark green models are inverted using picks with and without earthquake combined. The solid black curve is obtained without any picks. In the shallow section, inverted models from DAS data agree very well with the SASW survey, and both show the velocity inversion at a depth of ~15 m. $V_S$ inverted from car pickings (dark green) can be unreliable below ~100 m, because of the lack of low frequencies below 2.7 Hz.

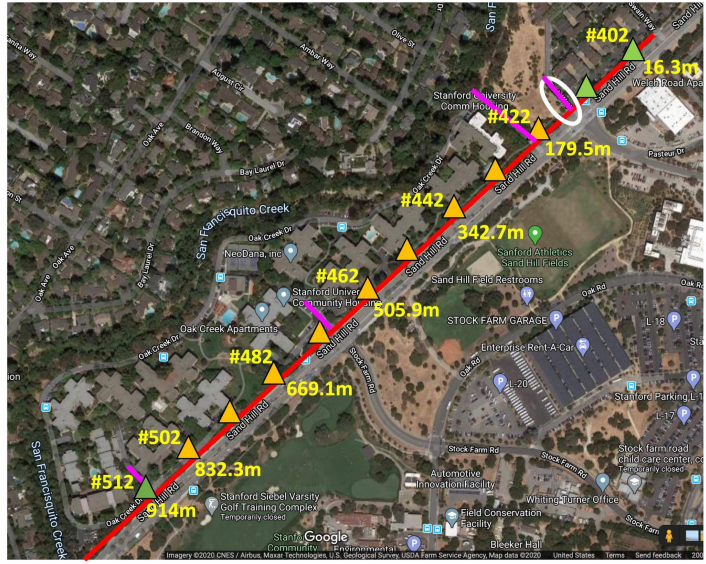

Figure 5: Satellite map of the Sandhill fiber (red line). Locations where both Vs and Poisson's ratio inversion are performed are shown with orange triangles, whereas, Poisson's ratio-only locations are shown with green triangles. Magenta lines indicate intersections. The white ellipse circles out the area where we conducted fiber location experiments by driving our car orthogonal to the fiber.

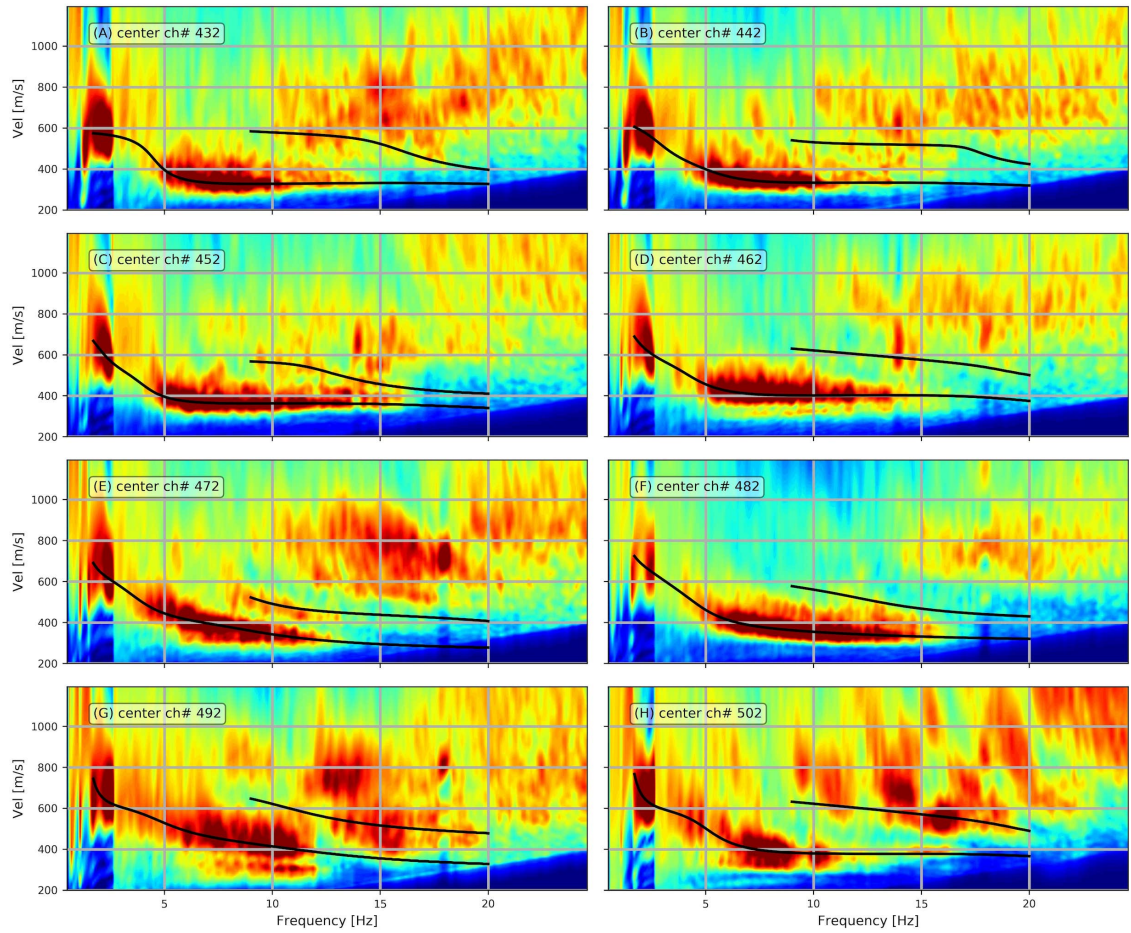

Figure 6: Dispersion images for eight locations (indicated in Figure 5) along the fiber. Frequencies below 2.7 Hz are from the Portola Valley earthquake. Above 2.7 Hz are from 33-car stack dispersion images. The proposed approach that does not require dispersion curve pickings are used for Vs inversion. The simulated dispersion curves using the inverted model are overlaid on the dispersion images using black curves. Inverted Vs models are shown in Figure 7.

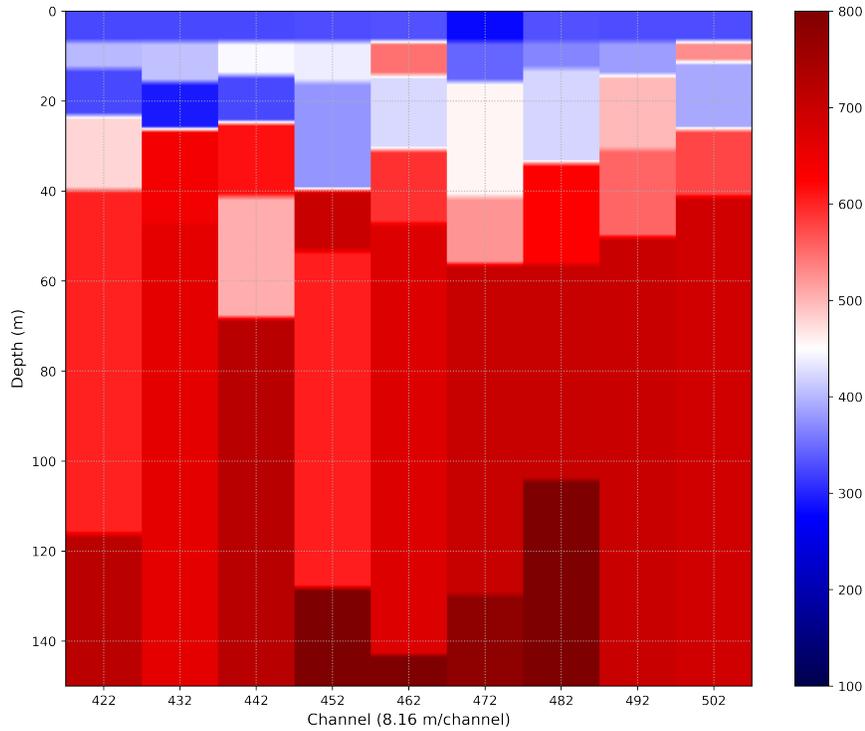

Figure 7: Inverted Vs models. The profile of the channel 422 comes from the solid black curves in Figure 4 (b). Results of the other channels are inverted from dispersion images shown in Figure 6. Except channel 472 and 492, we have velocity inversion at ~15 m. The top 25 m possible to be a deposit layer tends to have velocity below 500 m/s. Layers below 45 m with Vs above 600 m/s could be interpreted as hard rock.

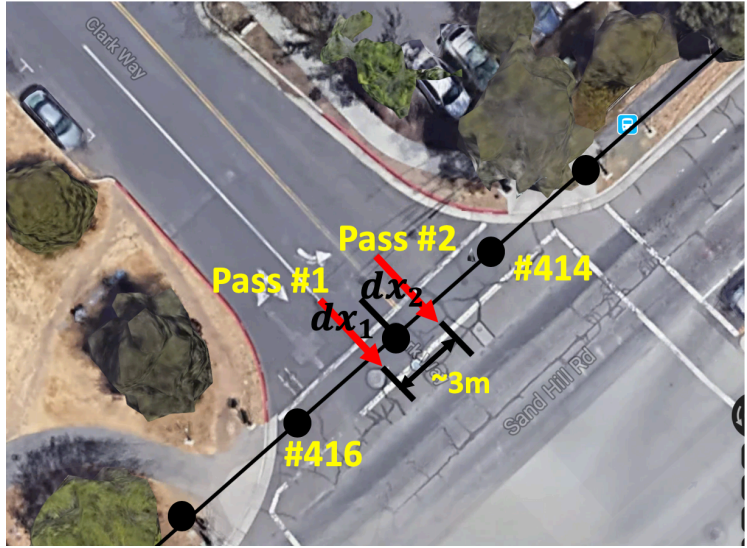

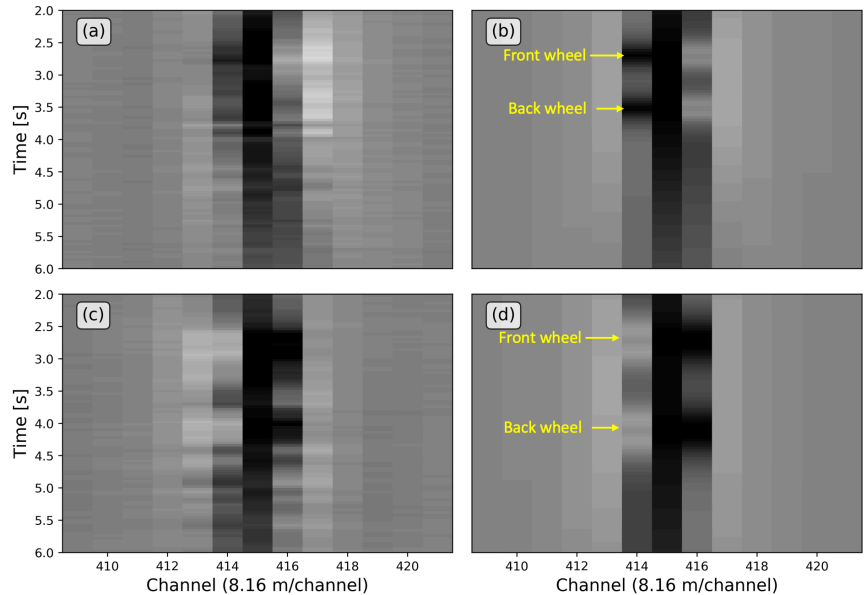

Figure 8: (Top) A zoomed-in map view of the intersection for our car-passing experiment. Car speeds for the two passes are respectively 3.4 m/s and 2 m/s. (Bottom) (a) and (c) show the data produced by the pass #1 and #2, respectively. Figure (b) and (d) show the simulated data using the parameters inverted through grid searches. The normalized cross-correlations are 0.94 between (a) and (b), and 0.96 between (c) and (d). The parameters of the best fit for the first pass are dx1=1.12 m, z=0.6 m and $\nu = 0.33$. For the second pass are dx2=-1.88 m, z=0.6 m, and $\nu = 0.35$. These results agree with the fact that the two passes are 3 m apart. The two inversions lead to a same fiber depth, 0.6 m. their Poisson's ratio estimates are close. The average estimated $\nu$ is 0.34 agreeing with a pavement material, Asphalt Concrete of a typical Poisson's ratio of 0.35 ($\pm$). Imprints of front and back wheels are pointed out with arrows in (b) and (d).

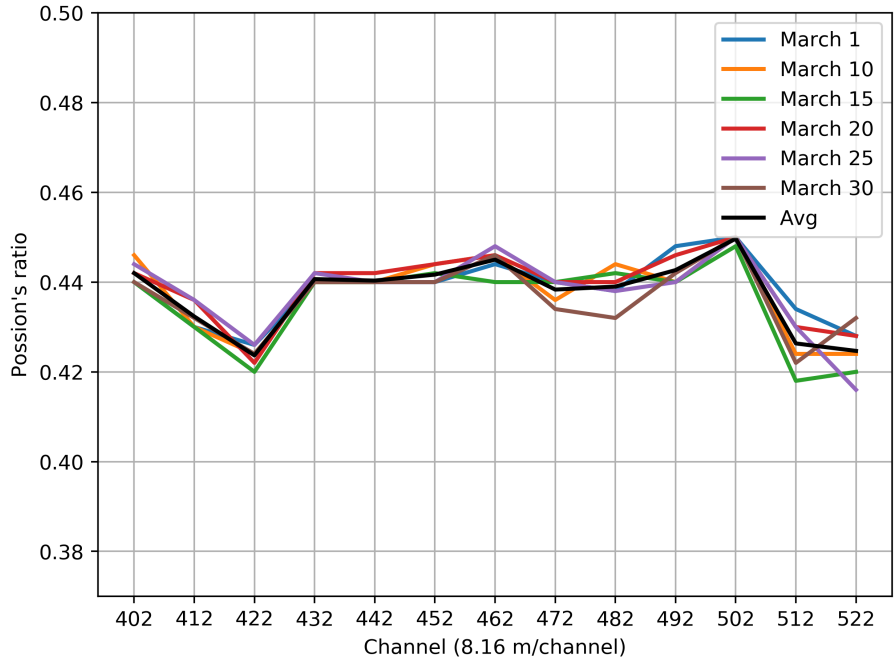

Figure 9: Daily averages of the Poisson's ratio estimates for the six days in March 2020. Poisson's ratios on average are ~0.44, indicating the shallow subsurface to be saturated clay. see Poisson's ratio drops at channels 422, 472 and 512 are likely related to their close locations to intersections, where fiber is buried under the road.